\documentclass[12pt]{article}
\makeatletter
\newcommand{\singlespacing}{\let\CS=\@currsize\renewcommand{\baselinestretch}{1}\tiny\CS}
\usepackage{epsfig}
\usepackage{amsbsy}
\usepackage{amssymb,color}
\begin{document}
\baselineskip=24pt
%\singlespacing
%\doublespacing
\parskip = 10pt
\def \qed {\hfill \vrule height7pt width 5pt depth 0pt}
\newcommand{\ve}[1]{\mbox{\boldmath$#1$}}
\newcommand{\IR}{\mbox{$I\!\!R$}}
\newcommand{\1}{\Rightarrow}
\newcommand{\bs}{\baselineskip}
\newcommand{\esp}{\end{sloppypar}}
\newcommand{\be}{\begin{equation}}
\newcommand{\ee}{\end{equation}}
\newcommand{\beanno}{\begin{eqnarray*}}
\newcommand{\inp}[2]{\left( {#1} ,\,{#2} \right)}
\newcommand{\eeanno}{\end{eqnarray*}}
\newcommand{\bea}{\begin{eqnarray}}
\newcommand{\eea}{\end{eqnarray}}
\newcommand{\ba}{\begin{array}}
\newcommand{\ea}{\end{array}}
\newcommand{\nno}{\nonumber}
\newcommand{\dou}{\partial}
\newcommand{\bc}{\begin{center}}
\newcommand{\ec}{\end{center}}
\newcommand{\2}{\subseteq}
\newcommand{\cl}{\centerline}
\newcommand{\ds}{\displaystyle}
\def\refhg{\hangindent=20pt\hangafter=1}
\def\refmark{\par\vskip 2.50mm\noindent\refhg}

\title{\sc Bivariate Discrete Generalized Exponential Distribution}

\author{Vahid Nekoukhou$^{1}$ \& Debasis Kundu$^{2}$ }

\date{}
\maketitle

\begin{abstract}
 In this paper we develop a bivariate discrete generalized exponential distribution, whose marginals are discrete generalized
exponential distribution as proposed by Nekoukhou, Alamatsaz and Bidram (``Discrete generalized exponential distribution of a
second type'', Statistics, 47, 876 - 887, 2013).  It is observed that
the proposed bivariate distribution is a very flexible distribution and the bivariate geometric  distribution can be obtained as a
special case of this distribution.  The proposed distribution can be seen as a natural discrete analogue of the bivariate generalized exponential
distribution proposed by Kundu and Gupta (``Bivariate generalized exponential distribution'', Journal of Multivariate Analysis,
100, 581 - 593, 2009).  We study different properties
of this distribution and explore its dependence structures.  We propose a new
EM algorithm to compute the maximum likelihood estimators of the unknown parameters which can be implemented very
efficiently, and discuss some inferential issues
also.  The analysis of one data set has been performed to show the effectiveness of the proposed model.
Finally we propose some open problems and conclude the paper.

\end{abstract}



\noindent$^1$Department of Statistics, Khansar Faculty of Mathematics and Computer Science, Khansar, Iran.

\noindent$^2$Department of Mathematics and Statistics, Indian Institute of Technology Kanpur, Kanpur,
Pin 208016, India.  \ \ e-mail: kundu@iitk.ac.in.

\newpage

\section{\sc Introduction}

The Generalized exponential (GE) distribution originally introduced by Gupta and Kundu \cite{GK:1999} has received considerable
attention in recent years.  It is an absolutely continuous univariate distribution with several interesting properties.  It
has been used quite successfully as an alternative to a gamma or a Weibull distribution.
Recently, Nekoukhou et al. \cite{NAB:2013} introduced a discrete generalized exponential (DGE) distribution, which can be considered as
the discrete analogue of the absolutely continuous GE distribution of Gupta and Kundu \cite{GK:1999}.  The DGE distribution proposed by
Nekoukhou et al. \cite{NAB:2013} is a very flexible two-parameter distribution.  The probability mass function of the DGE distribution
can be a decreasing or a unimodal function.  Similarly, the hazard function of the DGE distribution can be an increasing, decreasing
or a constant function depending
on the shape parameter.  Hence, the geometric distribution can be obtained as a special case of the DGE distribution.  It has been used
to analyze various discrete data sets, and the performances are quite satisfactory.

Discrete bivariate data also occur quite naturally in practice.  For example, the number of goals scored by two competing teams
or the number of insurance claims for two different causes is purely discrete in nature.  Recently, Lee and Cha \cite{LC:2015}
proposed two very general methods namely (i) minimization and (ii) maximization methods,
to generate a class of discrete bivariate distributions.  They have discussed in details some special cases namely
bivariate Poisson, bivariate geometric, bivariate negative binomial and bivariate binomial distributions.  Although, the
method proposed by Lee and Cha \cite{LC:2015} can produce a very flexible class of discrete bivariate distributions, the joint
probability mass function (PMF) may not be in a very convenient form in many cases.  Due to this reason, developing the inference
procedure of the unknown parameters becomes difficult in many cases.  Another point may be mentioned that the bivariate discrete
distributions proposed by Lee and Cha \cite{LC:2015} may not have the same corresponding univariate marginals.  For example,
the bivariate discrete Poisson distribution proposed by them does not have Poisson marginals, which may not be very
desirable.

The main aim of this paper is to consider the bivariate discrete generalized exponential (BDGE) distribution which can be obtained from
three independent DGE distributions by using the maximization method as suggested by Lee and Cha \cite{LC:2015}.  It is observed that
the BDGE distribution
is a natural discrete analogue of the bivariate generalized exponential distribution (BGE) proposed by Kundu and Gupta \cite{KG:2009}.
The BDGE distribution is a very flexible bivariate distribution, and its joint PMF can take various shapes depending on the
parameter values.  The generation from a BDGE distribution is straight forward, hence the simulation experiments can be
performed quite conveniently.  It has some interesting physical interpretations also.  In addition, its marginals are DGE distributions and the bivariate geometric distribution can be obtained as a special case of this model.

The estimation and the construction of confidence intervals of the unknown parameters play
an important role in any statistical problem.  The BDGE model has four parameters.  The maximum likelihood estimators (MLEs) of the
unknown parameters cannot be obtained in explicit forms.  They can be obtained by solving a four dimensional optimization
problem.  Algorithms like Newton-Raphson or Gauss-Newton method may be used to solve this problem.  But they have the standard problem
of convergence to a local optimum rather than the global optimum and choice of the initial guesses.  To avoid those problems, we propose to
use an EM algorithm to compute the MLEs of the unknown parameters.  We treat this problem as a missing value problem.
In each E-Step we use the maximum likelihood predictor method to estimate the missing values and it avoids computation of the
explicit expectation.  At each M-step
of the EM algorithm, the maximization of the 'pseudo' log-likelihood function can be performed by solving one non-linear equation only.
Hence, the implementation of the EM algorithm is quite simple in practice.  Moreover, at the last step of the EM algorithm using the idea of
Louis \cite{Louis:1982} the observed Fisher information matrix also can be obtained, and it will be used for construction of the
confidence intervals
of the unknown parameters.  One real data set has been analyzed to see the effectiveness of
the proposed model and the EM algorithm.  The performances are quite satisfactory.  Finally we propose some open problems.

Rest of the paper is organized as follows.  In Section 2 we provide the preliminaries.  The BDGE distribution is proposed and its
properties are discussed in Section 3.  In Section 4, we provide statistical inference procedures of the unknown parameters of a
BDGE model.  In Section 5 we provide the analysis of a real data set.  Finally, in Section 6 we propose some
open problems and conclude the paper.

\section{\sc Preliminaries}

\subsection{\sc The DGE Distribution}

The absolutely continuous GE distribution was proposed by Gupta and Kundu \cite{GK:1999} as an alternative to the well known gamma and Weibull distributions.  The
two-parameter GE distribution has the following probability density function (PDF) and cumulative distribution function (CDF), respectively;
\bea
f_{GE}(x; \alpha, \lambda) & = & \alpha \lambda e^{-\lambda x} (1 - e^{-\lambda x})^{\alpha-1}; \ \ \ x > 0,  \label{pdf-ge}  \\
F_{GE}(x; \alpha, \lambda) & = & (1 - e^{-\lambda x})^{\alpha}; \ \ \ x > 0.   \label{cdf-ge}
\eea
Here $\alpha > 0$ and $\lambda > 0$ are the shape and the scale parameters, respectively.  From now on a GE distribution with the shape parameter $\alpha$
and the scale parameter $\lambda$ will be denoted by GE$(\alpha,\lambda)$.

Recently, the DGE distribution was proposed by Nekoukhou et al. \cite{NAB:2013}.  It has been defined as follows.  A discrete random variable $X$ is said to have a DGE
distribution with parameters $\alpha$ and $\ds p$ $\ds (= e^{-\lambda})$, if the probability mass function (PMF) of $X$ can be written as follows:
\be
f_{DGE}(x; \alpha,p) = P(X = x) = (1 - p^{x+1})^{\alpha} - (1-p^x)^{\alpha}; \ \ \ \hbox{for} \ \ \ x \in {\cal N}_0 = \{0, 1, 2, \ldots, \}.
\label{dge-pmf}
\ee
The corresponding CDF becomes
\be
F_{DGE}(x; \alpha, p) = P(X \le x) = \left \{ \matrix{0 & \hbox{if} & x < 0  \cr (1 - p^{[x]+1})^{\alpha} & \hbox{if} & x \ge 0.} \right .
   \label{dge-cdf}
\ee
Here $[x]$ denotes the largest integer less than or equal to $x$.  From now on a DGE distribution with parameters $\alpha$ and $p$
will be denoted by DGE$(\alpha,p)$.  The PMF and the hazard function (HF) of a DGE distribution can take various shapes.
The PMF can be a decreasing or a unimodal function, and the HF can be an increasing or a decreasing function.
A DGE model is appropriate for modelling both over and under-dispersed data, since, in this model, the variance can be
larger or smaller than the mean which is not the case for some of the standard classical discrete distributions.  The authors did not provide the
probability generating function (PGF) of DGE, which is quite useful for a discrete distribution.  We provide the PGF of DGE$(\alpha,p)$
for completeness purposes.  Let us consider the PGF of $\ds G_X(z) = E(z^X)$, for $|z| < 1$.  Therefore,
\beanno
G_X(z) & = & E(z^X) = \sum_{i=0}^{\infty} \left \{(1-p^{i+1})^{\alpha} - (1-p^i)^{\alpha} \right \}z^i  \\
& = & \sum_{i=0}^{\infty} \sum_{j=1}^{\infty} (-1)^j {\alpha\choose{j}}p^{ij} (1-p^j)z^i
= \sum_{j=1}^{\infty} (-1)^j {\alpha\choose{j}} \frac{1-p^j}{1-zp^j},
\eeanno
here $\ds {\alpha\choose{j}} = \alpha(\alpha-1) \ldots (\alpha-j+1)/j!$.  Note that to compute $G_X(z)$, we have used the
identity for $i = 0, 1, \ldots,$
\be
P(X = i) = (1-p^{i+1})^{\alpha} - (1-p^i)^{\alpha} = \sum_{j=1}^{\infty}(-1)^{j+1} {\alpha\choose{j}} p^{ji}(1-p^j).
\label{identity}
\ee

The following representation of a DGE random variable becomes very useful.  Suppose $X \sim DGE(\alpha,p)$, then for
$\ds \lambda = -\ln p$,
\be
Y \sim \hbox{GE}(\alpha,\lambda) \Longrightarrow X = [Y] \sim \hbox{DGE}(\alpha, p).    \label{repre}
\ee
Using (\ref{repre}), the generation of a random sample from a DGE$(\alpha,p)$ becomes very simple.  For example, first we can generate
a random sample $Y$ from a GE$(\alpha,\lambda)$, and then considering $X = [Y]$, we can obtain a generated sample from
 DGE$(\alpha,p)$.   Suppose $Y_1 \sim$ GE$(\alpha_1,\lambda)$, $Y_2 \sim$ GE$(\alpha_2,\lambda)$, $X_1 = [Y_1]$, $X_2 = [Y_2]$, then
using (\ref{repre}) the following results can be easily obtained.

\noindent {RESULT 1:} If $Y_1$ and $Y_2$ are independently distributed then
$$
P(X_1 < X_2 ) = \sum_{j=0}^{\infty} \left \{ (1-p^{j+2})^{\alpha_2} - (1-p^{j+1})^{\alpha_2} \right \} (1-p^{j+1})^{\alpha_1}
\le \frac{\alpha_2}{\alpha_1+\alpha_2}.
$$
\noindent {\sc Proof:}  The first part (equality) of the above result follows from the definition as given below:
\beanno
P(X_1 < X_2) &  =  & \sum_{j=0}^{\infty} P(X_2 = j+1)P(X_1 \le j) \\
& = & \sum_{j=0}^{\infty} \left \{ (1-p^{j+2})^{\alpha_2} - (1-p^{j+1})^{\alpha_2} \right \} (1-p^{j+1})^{\alpha_1}.
\eeanno
The second part follows from the following observation:  Since $X_1 < X_2$ implies $Y_1 < Y_2$, therefore we have
$$
P(X_1 < X_2) \le P(Y_1 < Y_2) = \frac{\alpha_2}{\alpha_1+\alpha_2}.
$$
\qed

\noindent {RESULT 2:} If $X_1 \sim$ DGE$(\alpha_1, p)$, $X_2 \sim$ DGE$(\alpha_2,p)$ and $\alpha_1 < \alpha_2$, then
$X_2 >_{st} X_1$, i.e. $X_2$ is stochastically larger than $X_1$.

\noindent {\sc Proof:} Obvious. \qed

\noindent {RESULT 3:} If $X_1 \sim$ DGE$(\alpha_1,p)$, $\ldots$, $X_n \sim$ DGE$(\alpha_n,p)$, and they are independently
distributed then $\max\{X_1, \ldots, X_n\} \sim$ DGE$\ds \left (\sum_{i=1}^n \alpha_i,p \right )$.

\noindent {\sc Proof:} It follows from the CDF of the DGE distribution. \qed

\noindent {RESULT 4:} Let $Y \sim$ GE$(\alpha,\lambda)$ with $\lambda=-\ln p$, $X = [Y]$ and $U = Y - X$.  Then the conditional PDF of $U$ given $X = j$, for $\lambda=-\ln p$ and $j = 1, 2, \ldots$, is
$$
f_{U|X=j}(u) = \frac{\alpha \lambda (1-p^{j+u})^{\alpha-1} p^{j+u}}{(1-p^{j+1})^{\alpha} - (1-p^j)^{\alpha}}; \ \ \  0 < u < 1,
$$
and the PDF of $U$ is
$$
f_U(u) = \sum_{j=1}^{\infty} \alpha \lambda (1-p^{j+u})^{\alpha-1} p^{j+u}; \ \ \  0 < u < 1.
$$
\noindent{\sc Proof:} It is simple, hence, the details are avoided.   \qed

\subsection{\sc The BGE Distribution}

Kundu and Gupta \cite{KG:2009} introduced the bivariate generalized exponential (BGE) distribution, whose marginals are
GE distributions.  The model has also some interesting physical interpretations. The
joint CDF of the BGE model is as follows:
\be
F_{BGE}(y_1,y_2) = \left \{ \matrix{F_{GE}(y_1: \alpha_1 + \alpha_3, \lambda) F_{GE}(y_2; \alpha_2, \lambda) &
\hbox{if} & y_1 < y_2  \cr
F_{GE}(y_1; \alpha_1, \lambda) F_{GE}(y_2: \alpha_2 + \alpha_3, \lambda) &
\hbox{if} & y_1 > y_2  \cr
F_{GE}(y_1: \alpha_1+\alpha_2+\alpha_3, \lambda) & \hbox{if} & y_1 = y_2. \cr}
   \right .
\ee
The corresponding joint PDF becomes:
\be
f_{BGE}(y_1,y_2) = \left \{\matrix{f_1(y_1,y_2) & \hbox{if} & y_1 < y_2  \cr
f_2(y_1,y_2) & \hbox{if} & y_1 > y_2  \cr
f_0(y) & \hbox{if} & y_1 = y_2 = y,  \cr} \right .
\ee
where
\beanno
f_1(y_1,y_2) & = & f_{GE}(y_1; \alpha_1+\alpha_3, \lambda) f_{GE}(y_2; \alpha_2, \lambda),  \\
f_2(y_1,y_2) & = & f_{GE}(y_1; \alpha_1, \lambda) f_{GE}(y_2; \alpha_2+\alpha_3, \lambda),  \\
f_0(y) & = & \frac{\alpha_3}{\alpha_1+\alpha_2+\alpha_3}f_{GE}(y; \alpha_1+\alpha_2+\alpha_3, \lambda).
\eeanno
Kundu and Gupta \cite{KG:2009} provided several properties and discussed inferential issues of the above
mentioned model in details.  For some recent work on the BGE distribution one is referred to Ashour et al. \cite{AAM:2009},
Dey and Kundu \cite{DK:2012}, Dewan and Nandi \cite{DN:2013}, Genc \cite{Genc:2014} and the references cited there in.

\section{\sc The BDGE Distribution and its Properties}

\subsection{\sc Definition and Interpretations}

\noindent {\sc Definition:}  Suppose $U_1\sim$ DGE$(\alpha_1,p)$,  $U_2 \sim$ DGE$(\alpha_2,p)$ and $U_3 \sim$ DGE$(\alpha_3,p)$ and
they are independently distributed. If $X_1=\max\{U_1,U_3\}$ and $X_2=\max\{U_2,U_3\}$, then we say that the bivariate vector
$(X_1,X_2)$ has a BDGE distribution with the parameter vector
$\boldsymbol{\theta}=(\alpha_1,\alpha_2,\alpha_3,p)^T$.  From now on we will denote this discrete bivariate distribution by
BDGE$(\alpha_1,\alpha_2,\alpha_3,p)$.

\noindent If $(X_1,X_2)\sim$ BDGE$(\alpha_1,\alpha_2,\alpha_3,p)$, then the joint CDF of $(X_1,X_2)$ for $x_1 \in {\cal N}_0$,
$x_2 \in {\cal N}_0$ and for $z=\min\{x_1,x_2\}$ is
\begin{eqnarray}\label{r6}
F_{X_1,X_2}(x_1,x_2)&=&(1-p^{x_1+1})^{\alpha_1}(1-p^{x_2+1})^{\alpha_2}(1-p^{z+1})^{\alpha_3}   \nonumber  \\
&=&F_{DGE}(x_1; \alpha_1, p)F_{DGE}(x_2; \alpha_2, p)F_{DGE}(z; \alpha_3, p),  \nonumber  \\
& = & \left\{%
\begin{array}{ll}
F_{DGE}(x_1; \alpha_1+\alpha_3, p)F_{DGE}(x_2; \alpha_2, p) & \hbox{if~~~~ $x_1<x_2$} \\
F_{DGE}(x_1; \alpha_1)F_{DGE}(x_2; \alpha_2+\alpha_3, p) & \hbox{if~~~~ $x_2<x_1$} \\
F_{DGE}(x; \alpha_1+\alpha_2+\alpha_3, p) & \hbox{if~~~~$x_1=x_2=x$}.
\end{array}%
\right.
\end{eqnarray}
The corresponding joint PMF of $(X_1,X_2)$ for $x_1, x_2 \in {\cal N}_0$ is given by
\begin{eqnarray}\label{r8}\nonumber
f_{X_1,X_2}(x_1,x_2)=\left\{%
\begin{array}{ll}
f_1(x_1,x_2) & \hbox{if~~~~ $0\leq x_1<x_2$} \\
f_2(x_1,x_2) & \hbox{if~~~~ $0\leq x_2<x_1$} \\
f_0 (x)& \hbox{if~~~~$0\leq x_1=x_2=x$},\\
\end{array}%
\right.
\end{eqnarray}
where
$$f_1(x_1,x_2)=f_{DGE}(x_1; \alpha_1+\alpha_3, p)f_{DGE}(x_2; \alpha_2, p),$$
$$f_2(x_1,x_2)=f_{DGE}(x_1; \alpha_1, p)f_{DGE}(x_2; \alpha_2+\alpha_3, p),$$
$$f_0(x)= p_1 f_{DGE}(x; \alpha_1+\alpha_3, p)- p_2 f_{DGE}(x; \alpha_1, p),$$
and $\ds p_1=(1-p^{x+1})^{\alpha_2}$, $\ds p_2=(1-p^{x})^{\alpha_2+\alpha_3}$.  Note that
the expressions $f_1(x_1, x_2)$, $f_2(x_1, x_2)$ and $f_3(x_1, x_2)$ for $x_1, x_2 \in {\cal N}_0$ can be easily obtained by using
the relation
$$
f_{X_1,X_2}(x_1,x_2) =
F_{X_1,X_2}(x_1,x_2) -F_{X_1,X_2}(x_1-1,x_2) - F_{X_1,X_2}(x_1,x_2-1) +
F_{X_1,X_2}(x_1-1,x_2-1).
$$
If $(X_1, X_2) \sim$ BDGE$(\alpha_1, \alpha_2, \alpha_3, p)$, then the joint survival function (SF) of the vector $(X_1,X_2)$ can also
be expressed in a compact form due to the relation
$$
S_{X_1,X_2}(x_1,x_2)=1-F_{X_1}(x_1)-F_{X_2}(x_2)+F_{X_1,X_2}(x_1,x_2).
$$
Now we provide the joint PGF of $X_1$ and $X_2$.  The joint PGF of $X_1$ and $X_2$ for $|z_1| < 1$ and $|z_2| < 1$, can be
written as
\beanno
G_{X_1,X_2}(z_1, z_2) & = & E(z_1^{X_1} z_2^{X_2}) = \sum_{j=0}^{\infty} \sum_{i=0}^{\infty} P(X_1 = i, X_2 = j) z_1^i z_2^j  \\
& = & \sum_{j=0}^{\infty} \sum_{i=0}^{j-1} \sum_{k=1}^{\infty} \sum_{l=1}^{\infty} (-1)^{k+l} {{\alpha_1+\alpha_3}\choose{k}}
{\alpha_2\choose{l}} p^{ki+lj} (1-p^k)(1-p^l) z_1^i z_2^j + \\
&  & \sum_{j=0}^{\infty} \sum_{i=j+1}^{\infty} \sum_{k=1}^{\infty} \sum_{l=1}^{\infty} (-1)^{k+l} {{\alpha_1}\choose{k}}
{{\alpha_2+\alpha_3}\choose{l}} p^{ki+lj} (1-p^k)(1-p^l) z_1^i z_2^j + \\
&  & \sum_{j=0}^{\infty} \sum_{i=0}^{\infty} \sum_{k=1}^{\infty} (-1)^{j+k+1} {{\alpha_2}\choose{j}}
{{\alpha_1+\alpha_3}\choose{k}} p^{ki+ji+1} (1-p^k)(1-p^l) z_1^i z_2^i - \\
&  & \sum_{j=0}^{\infty} \sum_{i=0}^{\infty} \sum_{k=1}^{\infty} (-1)^{j+k+1} {{\alpha_2+\alpha_3}\choose{j}}
{{\alpha_1}\choose{k}} p^{ki+ji} (1-p^k)(1-p^l) z_1^i z_2^i.
\eeanno
Using the joint PGF, different moments and product moments can be obtained as infinite series.

\noindent The following shock model and maintenance model interpretations can be provided for the BDGE distribution.

\noindent {\sc Shock Model:} Suppose a system has two components, and it is assumed that the amount of shocks is measured in a
digital (discrete) unit.  Each component is subjected to individual shocks
say $U_1$ and $U_2$, respectively.  The system faces an overall shock $U_3$, which is transmitted to both the component
equally, independent of their individual shocks.  Therefore, the observed shocks at the two components are $X_1 = \max\{U_1, U_3\}$
and $X_2  = \max\{U_2, U_3\}$.

\noindent {\sc Maintenance Model:}  Suppose a system has two components and it is assumed that each component has been
maintained independently and also there is an overall maintenance.  Due to component maintenance, suppose the
lifetime of the individual component is increased by $U_i$ amount and because of the overall maintenance, the
lifetime of each component is increased by $U_3$ amount.  Here, $U_1$, $U_2$ and $U_3$ are all measured in a discrete unit.
Therefore, the  increased lifetimes of the two components
are $X_1 = \max\{U_1, U_3\}$ and $X_2 = \max\{U_2, U_3\}$, respectively.

\subsection{\sc Properties}

\noindent {RESULT 5:} If $(Y_1, Y_2) \sim$ BGE$(\alpha_1, \alpha_2, \alpha_3, \lambda)$, then $(X_1, X_2) \sim$ BDGE$(\alpha_1,
\alpha_2, \alpha_3, p)$, where $X_1 = [Y_1]$, $X_2 = [Y_2]$ and $p=e^{-\lambda}$.

\noindent {\sc Proof:} It can be easily obtained from the joint CDF of $X_1$ and $X_2$.  \qed

The Result 5 indicates that the proposed BDGE distribution is a natural discrete version of BGE distribution.  In addition, the marginals are DGE distributions. More precisely, we see that $X_1 \sim$ DGE$(\alpha_1+\alpha_3,p)$ and $X_2 \sim$ DGE$(\alpha_2+\alpha_3,p)$.
The following
algorithm can be used to generate a random sample from a BDGE distribution using Result 4.

\noindent {\sc Algorithm:}

\begin{itemize}

\item Generate $U_1 \sim$ GE$(\alpha_1, \lambda)$, $U_2 \sim$ GE$(\alpha_2,\lambda)$, $U_3 \sim$ GE($\alpha_3,\lambda$)
by using inverse transformation method.

\item Obtain $Y_1 = \max\{U_1, U_3\}$ and $Y_2 = \max\{U_2, U_3\}$.

\item $(X_1, X_2)$, where $X_1 = [Y_1]$ and $X_2 = [Y_2]$, is the desired random sample.
\end{itemize}

\noindent {RESULT 6:}  We have the following results regarding the conditional distribution of $X_1$ given $X_2$
when $(X_1,X_2)\sim$ BDGE$(\alpha_1,\alpha_2,\alpha_3,p)$.  The proofs are quite standard and the details are avoided.

\noindent (a) The conditional PMF of $X_1$ given $X_2=x_2$, say $f_{X_1|X_2=x_2}(x_1|x_2)$, is given by
\begin{eqnarray}\label{r80}\nonumber
f_{X_1|X_2=x_2}(x_1|x_2)=\left\{%
\begin{array}{ll}
f_1(x_1|x_2) & \hbox{if~~~~ $0\leq x_1<x_2$} \\
f_2(x_1|x_2) & \hbox{if~~~~ $0\leq x_2<x_1$} \\
f_0 (x_1|x_2)& \hbox{if~~~~$0\leq x_1=x_2=x$},\\
\end{array}%
\right.
\end{eqnarray}
where $$f_i(x_1|x_2)=\frac{f_i(x_1,x_2)}{f_{DGE}(x_2;\alpha_2+\alpha_3)}, ~i=1,2$$
and
$$f_0(x_1|x_2)=\frac{f_0(x)}{f_{DGE}(x_2;\alpha_2+\alpha_3)}.$$

\noindent (b) The conditional CDF of $X_1$ given $X_2\leq x_2$, say $F_{X_1|X_2\leq x_2}(x_1)$, is given by
\begin{eqnarray}\label{r9}\nonumber
F_{X_1|X_2\leq x_2}(x_1)&=&P(X_1\leq x_1|X_2\leq x_2)\\\nonumber
&=&\left\{%
\begin{array}{ll}
(1-p^{x_1+1})^{\alpha_1+\alpha_3}(1-p^{x_2+1})^{-\alpha_3} & \hbox{if~~~~ $0\leq x_1<x_2$} \\
(1-p^{x_1+1})^{\alpha_1} & \hbox{if~~~~ $0\leq x_2<x_1$} \\
(1-p^{x+1})^{\alpha_1}& \hbox{if~~~~$0\leq x_1=x_2=x$}.
\end{array}%
\right.
\end{eqnarray}
(c) The conditional CDF of $X_1$ given $X_2=x_2$, say $F_{X_1|X_2=x_2}(x_1)$, is given by
\begin{eqnarray}\label{r10}\nonumber
F_{X_1|X_2=x_2}(x_1)&=&P(X_1\leq x_1|X_2=x_2)\\\nonumber
&=&\left\{%
\begin{array}{ll}
\frac{F_{DGE}(x_1;\alpha_1+\alpha_3)f_{DGE}(x_2;\alpha_2)}{f_{DGE}(x_2;\alpha_2+\alpha_3)} & \hbox{if~~ $0\leq x_1<x_2$} \\
F_{DGE}(x_1;\alpha_1) & \hbox{if~~ $0\leq x_2<x_1$} \\
\frac{F_{DGE}(x;\alpha_1+\alpha_2+\alpha_3)-F_{DGE}(x;\alpha_1)F_{DGE}(x-1;\alpha_2+\alpha_3)}{f_{DGE}(x_2;\alpha_2+\alpha_3)} & \hbox{if~~$0\leq x_1=x_2=x.$}
\end{array}%
\right.
\end{eqnarray}

Note that if $\alpha_3=\alpha$, $0<\alpha<1$, and also $\alpha_1=\alpha_2=1-\alpha$, then $(X_1,X_2)$ has geometric marginals.  Therefore, we have a new bivariate geometric distribution with parameters \linebreak
$0 < p < 1$ and $0 < \alpha < 1$, and whose joint CDF is
\be
F_{X_1,X_2}(x_1,x_2) = (1-p^{x_1+1})^{1-\alpha}(1-p^{x_2+1})^{1-\alpha}(1-p^{z+1})^{\alpha}.
\ee
Here $x_1, x_2 \in {\cal N}_0$ and $z = \min\{x_1, x_2\}$ as before.  Moreover in this case $X_1$ and $X_2$ both have geometric
distributions with parameter $p$.

\noindent From (\ref{r6}) it follows that for all values of $x_1\geq 0$ and $x_2\geq 0$,
$$
F_{X_1,X_2}(x_1,x_2)\geq F_{X_1}(x_1)F_{X_2}(x_2).
$$
Therefore, $X_1$ and $X_2$ are positive quadrant dependent, i.e., for every pair of increasing functions $m_1(.)$ and $m_2(.)$, it follows that $Cov(m_1(X_1),m_2(X_2))\geq 0$, see for example Nelsen \cite{Nelsen:2006}. Moreover, in view of the fact that
$$
S_{X_1,X_2}(x_1,x_2)-S_{X_1}(x_1)S_{X_2}(x_2)=F_{X_1,X_2}(x_1,x_2)-F_{X_1}(x_1)F_{X_2}(x_2),
$$
we see that
$$
S_{X_1,X_2}(x_1,x_2)\geq S_{X_1}(x_1)S_{X_2}(x_2).
$$
Moreover, $X_1$ and $X_2$ are independent when $\alpha_3$ = 0.  Therefore, $Corr\{X_1, X_2\} = 0$.  For fixed $\alpha_1$ and
$\alpha_2$
$$
\lim_{\alpha_3 \rightarrow \infty} P(U_1 > U_3) = \lim_{\alpha_3 \rightarrow \infty} P(U_1 = U_3) =  \lim_{\alpha_3 \rightarrow \infty} P(U_2 > U_3) =
\lim_{\alpha_3 \rightarrow \infty} P(U_2 = U_3) = 0,
$$
where $U_1, U_2$ and $U_3$ are same as defined in Section 3.1. Now, for any two arbitrary random variables $Y_1$ and $Y_2$ with finite second moments, let us define a new random variable
$$g(Y_1, Y_2) = \frac{\{Y_1 - E(Y_1)\}\{Y_2 - E(Y_2)\}}{\sqrt{V(Y_1)}\sqrt{V(Y_2)}}.
$$
It is clear that $E|g(Y_1, Y_2)| \le 1$, for any two arbitrary random variables $Y_1$ and $Y_2$ with finite second moments.
Therefore,
\beanno
\lim_{\alpha_3 \rightarrow \infty} Corr\{X_1, X_2\}  & =  & \lim_{\alpha_3 \rightarrow \infty} E(g(X_1, X_2)) \\ & = &
\lim_{\alpha_3 \rightarrow \infty} E(g(U_3,U_3)) \times P(U_1 < U_3, U_2 < U_3) \\
&  = & \lim_{\alpha_3 \rightarrow \infty} P(U_1 <  U_3, U_2 <  U_3) = 1.
\eeanno
Therefore, $Corr\{X_1, X_2\} \rightarrow 1$.  Hence,
for a BDGE distribution the correlation coefficient has the range $[0, 1)$.

Let us recall the following two definitions.  Let $(X, Y)$ be a pair of random variables, then (a) $Y$ said to be left-tail
decreasing in $X$, if and only if $P(Y \le y|X \le x)$ is a non-increasing function of $x$ for every $y$, and (b)
$Y$ said to be stochastically increasing in $X$ if and only if $P(Y \le y|X = x)$ is a non-increasing function of $x$ for every $y$,
see Nelsen \cite{Nelsen:2006}.  We have the following result for a BDGE distribution.

\noindent {RESULT 7:} Suppose $(X_1,X_2)\sim$ BDGE$(\alpha_1,\alpha_2,\alpha_3,p)$, then $X_2$ is left-tail decreasing in $X_1$ and
$X_2$ is stochastically increasing in $X_1$.

\noindent {\sc Proof:} Both the assertions follow from Result 6.   \qed

Suppose $(X, Y)$ is a pair of discrete random variables having support on ${\cal N}_0 \times {\cal N}_0$, then it is said to have a total
positivity of order two (TP$_2$) property if the joint probability mass function $f(x,y)$ satisfies
\be
f(x_1,y_1) f(x_2, y_2) \ge f(x_2, y_1) f(x_1,y_2) \ \ \ \ \hbox{for all} \ \ \ \ x_1, y_1, x_2, y_2 \in {\cal N}_0.  \label{tp2}
\ee

\noindent {RESULT 8:} Suppose $(X_1,X_2)\sim$ BDGE$(\alpha_1,\alpha_2,\alpha_3,p)$, then it has the TP$_2$ property.

\noindent {\sc Proof:} Suppose $x_{11}, x_{21}, x_{12}, x_{22} \in {\cal N}_0$ and $x_{11} < x_{21} < x_{12} < x_{22}$, then observe that
$$
\frac{f(x_{11},x_{21}) f(x_{12}, x_{22})}{f(x_{12},x_{21}) f(x_{11}, x_{22})} = \left [ \frac{1-p^{x_{12}+1}}{1-p^{x_{21}+1}} \right ]^{\alpha_3} \ge 1.
$$
Similarly considering all other cases such as $x_{11} = x_{21} < x_{12} < x_{22}$, $x_{21} < x_{11} < x_{12} < x_{22}$ etc. it can be shown that
it satisfies (\ref{tp2}).  Hence, the result is proved.   \qed

The following result is an extension of Result 1 to the bivariate case.

\noindent {RESULT 9:} If $(X_1,X_2)\sim$ BDGE$(\alpha_1,\alpha_2,\alpha_3,p)$, then
$$
P(X_1 < X_2 ) = \sum_{j=0}^{\infty} \left \{ (1-p^{j+2})^{\alpha_2} - (1-p^{j+1})^{\alpha_2} \right \} (1-p^{j+1})^{\alpha_1+\alpha_3}
\le \frac{\alpha_2}{\alpha_1+\alpha_2+\alpha_3}.
$$
\noindent {\sc Proof:}  It follows exactly the same way as the proof of Result 1, hence the details are avoided.     \qed

The following simple result is useful for goodness of fitting purposes.  Suppose $(X_1,X_2) \sim$
BDGE$(\alpha_1,\alpha_2,\alpha_3,p)$, then $\max\{X_1, X_2\} \sim$ DGE$(\alpha_1+\alpha_2+\alpha_3, p)$.
The proof can be easily obtained as follows.
$$
P(\max\{X_1,X_2\}\leq x)=P(U_1\leq x, U_2\leq x, U_3\leq x)=(1-p^{x+1})^{\alpha_1+\alpha_2+\alpha_3}.
$$
In addition, the following result is an extension of Result 3 to the bivariate case.

\noindent {RESULT 10:} Suppose $(X_{i1}, X_{i2}) \sim$ BDGE$(\alpha_{i1}, \alpha_{i2}, \alpha_{i3}, p)$, for $i = 1, \ldots, n$,
and they are independently distributed.  If $\ds Y_1 = \max\{X_{11}, \ldots, X_{n1}\}$ and $\ds Y_2 = \max\{X_{12}, \ldots, X_{n2}\}$,
then $(Y_1, Y_2) \sim$ BDGE$\ds \left (\sum_{i=1}^n \alpha_{i1}, \sum_{i=1}^n \alpha_{i2}, \sum_{i=1}^n \alpha_{i3}, p \right )$.

\noindent {\sc Proof:} The proof can be easily obtained from the joint CDF and hence, the details are avoided. \qed

\section{\sc Statistical Inference}

\subsection{\sc Maximum Likelihood Estimation}

In this section we consider the maximum likelihood estimation of the unknown parameters of a BDGE$(\alpha_1, \alpha_2, \alpha_3, p)$
model based on a sample of size $n$.  It is assumed that we have the following bivariate sample ${\cal D } =
\{(x_{11}, x_{21}), \ldots, (x_{1n}, x_{2n})\}$.  We use the following notations
\be
I_1 = \{i: x_{1i} < x_{2i}\}, \ \ \ I_2 = \{i: x_{1i} > x_{2i}\}, \ \ \ \ I_0 = \{i: x_{1i} = x_{2i} = x_i\},   \label{data}
\ee
and
$\ds n_1 = |I_1|$, $\ds n_2 = |I_2|$, $\ds n_0 = |I_0|$.  Here $|I_j|$ denotes the number of elements in set $I_j$, for $j = 0, 1, 2$.
Based on the above data (\ref{data}), the log-likelihood function can be written as
\bea
l(\alpha_1, \alpha_2, \alpha_3, p| {\cal D}) & = & \sum_{i \in I_1} \ln \left [ (1 - p^{x_{1i}+1})^{\alpha_1+\alpha_3} -
(1 - p^{x_{1i}})^{\alpha_1+\alpha_3} \right ] + \nonumber  \\
&  & \sum_{i \in I_1} \ln \left [ (1 - p^{x_{2i}+1})^{\alpha_2} - (1 - p^{x_{2i}})^{\alpha_2} \right ] +   \nonumber  \\
&  & \sum_{i \in I_2} \ln \left [ (1 - p^{x_{1i}+1})^{\alpha_1} -
(1 - p^{x_{1i}})^{\alpha_1} \right ] + \nonumber  \\
&  & \sum_{i \in I_2} \ln \left [ (1 - p^{x_{2i}+1})^{\alpha_2+\alpha_3} - (1 - p^{x_{2i}})^{\alpha_2+\alpha_3} \right ] +  \nonumber \\
&  & \sum_{i \in I_0} \ln \left [ (1 - p^{x_i+1})^{\alpha_2} \left \{(1 - p^{x_i+1})^{\alpha_1+\alpha_3} - (1 - p^{x_i})^{\alpha_1+\alpha_3}  \right \}
\right .   -   \nonumber  \\
&  & \ \ \ \ \ \ \ \ \ \ \ \ \ \ \left . (1 - p^{x_i})^{\alpha_2+\alpha_3} \left \{(1 - p^{x_i+1})^{\alpha_1} - (1 - p^{x_i})^{\alpha_1}  \right \}
\right ].   \label{ll}
\eea
Therefore, the MLEs of the unknown parameters can be obtained by maximizing (\ref{ll}) with respect to the unknown parameters. It can
be obtained by solving four non-linear equations simultaneously.  Newton-Raphson method may be used to solve these four non-linear
equations.

We propose to use an EM algorithm technique to compute the MLEs of the
unknown parameters, which is very specific to a discrete distribution,  mainly to avoid solving four dimensional optimization problem.
We treat this problem as a missing
value problem as it is usually being done for any implementation of an EM algorithm.  We estimate the missing values by maximum likelihood
predictor method similar to Karlis \cite{Karlis:2003}.  The basic idea comes from the fact that if we
know $\{(u_{1i}, u_{2i}, u_{3i}); i = 1, \ldots, n\}$ then the log-likelihood function of the complete observations becomes
\be
l_{complete}(\alpha_1, \alpha_2, \alpha_3, p| {\cal D}) = g_1(\alpha_1,p) + g_2(\alpha_2,p) + g_3(\alpha_3,p),   \label{ll-com}
\ee
where
\bea
g_1(\alpha_1, p) & = & \sum_{i=1}^n \ln \left [ (1 - p^{u_{1i}+1})^{\alpha_1} - (1 - p^{u_{1i}})^{\alpha_1} \right ],   \label{g-1}  \\
g_2(\alpha_2, p) & = & \sum_{i=1}^n \ln \left [ (1 - p^{u_{2i}+1})^{\alpha_2} - (1 - p^{u_{2i}})^{\alpha_2} \right ],   \label {g-2} \\
g_3(\alpha_3, p) & = & \sum_{i=1}^n \ln \left [ (1 - p^{u_{3i}+1})^{\alpha_3} - (1 - p^{u_{3i}})^{\alpha_3} \right ].   \label{g-3}
\eea
The following result will be useful for further development.

\noindent {RESULT 11:} For any $0 < p < 1$, $g_1(\alpha_1,p)$, $g_2(\alpha_2,p)$ and $g_3(\alpha_3,p)$ as defined in (\ref{g-1}),
(\ref{g-2}) and (\ref{g-3}), respectively are unimodal functions of $\alpha_1$, $\alpha_2$ and $\alpha_3$, respectively.

\noindent {\sc Proof:} See in the Appendix. \qed

Therefore, the maximization of (\ref{ll-com}) can be easily performed using profile likelihood method.  For a fixed
$p$ maximize $g_1(\alpha_1, p)$, $g_2(\alpha_2, p)$ and $g_3(\alpha_3, p)$, with respect to $\alpha_1$, $\alpha_2$ and $\alpha_3$,
respectively, and they are unique due to Result 11.  If they are $\widehat{\alpha}_1(p)$,  $\widehat{\alpha}_2(p)$ and  $\widehat{\alpha}_3(p)$, respectively, then
obtain the MLE of $p$ as $\widehat{p}$ by maximizing
$$
g(p) = g_1(\widehat{\alpha}_1(p), p) + g_2(\widehat{\alpha}_2(p), p) + g_3(\widehat{\alpha}_3(p), p),
$$
and the MLEs of $\alpha_1$, $\alpha_2$ and $\alpha_3$ as
$$
\widehat{\alpha}_1 = \widehat{\alpha}_1(\widehat{p}), \ \ \ \widehat{\alpha}_2 = \widehat{\alpha}_2(\widehat{p}), \ \ \
\widehat{\alpha}_3 = \widehat{\alpha}_3(\widehat{p}).
$$
Hence, it is clear that if we have the complete observations, then the MLEs of the unknown parameters can be obtained by
solving three one dimensional optimization problems.

Therefore, to implement the EM algorithm at each E-Step, we obtain the missing $\{u_{ji}\}$ for $j = 1, 2, 3$, say $\{(\widetilde{u}_{1i},
\widetilde{u}_{2i}, \widetilde{u}_{3i}); i = 1, \ldots, n\}$, by using the maximum likelihood predictor method, and then at the corresponding
M-Step, we maximize the complete log-likelihood function to compute the estimates for the next iterate.  It
is assumed that at the $k$-th step the values of the unknown parameters are $\Theta^{(k)} = (\alpha_1^{(k)}, \alpha_2^{(k)}, \alpha_3^{(k)},
p^{(k)})$ and the available data are ${\cal D}^{(k)} = \{(\widetilde{u}_{1i}, \widetilde{u}_{2i}, \widetilde{u}_{3i}); i = 1, \ldots, n\}$.
We will provide the methodology how to compute $\Theta^{(k+1)}$ from $\Theta^{(k)}$ based on ${\cal D}^{(k)}$ by using EM algorithm.

\noindent E-Step:  We will mention how to obtain $(\widetilde{u}_{1i},\widetilde{u}_{2i}, \widetilde{u}_{3i})$ from $(x_{1i}, x_{2i})$ by maximum likelihood  prediction method as follows:

\noindent {\sc Case I:} \ $x_{1i} < x_{2i}$

  Since $x_{1i} = \max\{u_{1i}, u_{3i}\}$ and $x_{2i} = \max\{u_{2i}, u_{3i}\}$, it is clear $u_{1i} < u_{2i}$ and
$u_{3i} < u_{2i}$. Therefore, $\widetilde{u}_{2i} = x_{2i}$, and $(\widetilde{u}_{1i}, \widetilde{u}_{3i})$ are chosen so that it maximizes
the corresponding probability, i.e.
\be
(\widetilde{u}_{1i}, \widetilde{u}_{3i}) = \hbox{arg max}_{\{(u,v); \max\{u,v\} = x_{1i}\}}
f_{DGE}(u; \alpha_1^{(k)},p^{(k)})  f_{DGE}(v; \alpha_3^{(k)},p^{(k)}).
\ee

\noindent {\sc Case II:} \ $x_{2i} < x_{1i}$

In this case similarly as before, $\widetilde{u}_{1i} = x_{1i}$, and $(\widetilde{u}_{2i}, \widetilde{u}_{3i})$ are chosen so that
\be
(\widetilde{u}_{2i}, \widetilde{u}_{3i}) = \hbox{arg max}_{\{(u,v); \max\{u,v\} = x_{2i}\}}
f_{DGE}(u; \alpha_2^{(k)},p^{(k)}) f_{DGE}(v; \alpha_3^{(k)},p^{(k)}).
\ee

\noindent {\sc Case III:} \ $x_{1i} = x_{2i} = x_i$

In this case $(\widetilde{u}_{1i}, \widetilde{u}_{2i}, \widetilde{u}_{3i})$ are chosen so that
\be
(\widetilde{u}_{1i}, \widetilde{u}_{2i}, \widetilde{u}_{3i}) = \hbox{arg max}_{\{(u,v,w); \max\{u,v,w\} = x_i\}}
g(u,v,w; \alpha_1^{(k)}, \alpha_2^{(k)}, \alpha_3^{(k)},p^{(k)}),
\ee
where
$$
g(u,v,w; \alpha_1^{(k)}, \alpha_2^{(k)}, \alpha_3^{(k)},p^{(k)}) = f_{DGE}(u; \alpha_1^{(k)},p^{(k)})  f_{DGE}(v; \alpha_2^{(k)},p^{(k)})  f_{DGE}(w; \alpha_3^{(k)},p^{(k)}).
$$

\noindent M-Step:  In this step we maximize $l_{complete}(\alpha_1, \alpha_2, \alpha_3, p| {\cal D}^{(k)})$ as defined in (\ref{ll-com})
with respect to $\alpha_1, \alpha_2, \alpha_3$ and $p$ to obtain $\Theta^{(k+1)}$ by profile likelihood method as described before.

The choice of initial estimates of $\alpha_1$, $\alpha_2$, $\alpha_3$ and $p$ are important.  Based on the marginal $\{x_{1i}; i = 1,
\ldots, n\}$ we can obtain estimates of $\alpha_1+\alpha_3$ and $p$.  Similarly, from $\{x_{2i}; i = 1, \ldots, n\}$, we can obtain
estimates of $\alpha_2+\alpha_3$ and $p$, and from $\{z_i = \max\{x_{1i}, x_{2i}\}; i = 1, \ldots, n\}$, we can obtain initial estimates
of $\alpha_1+\alpha_2+\alpha_3$ and $p$.  From these estimates we can obtain initial estimates of $\alpha_1$, $\alpha_2$, $\alpha_3$ and
$p$.

Therefore, the EM algorithm can be implemented as follows:

\noindent {\sc Algorithm:}

\begin{enumerate}

\item Initial Estimate: Get initial estimates of $\alpha_1$, $\alpha_2$, $\alpha_3$ and $p$ from the marginals and from the maximum of the two
marginals.  Let it be denoted by $\Theta^{(0)} = (\alpha_1^{(0)}, \alpha_2^{(0)}, \alpha_3^{(0)}, p^{(0)})$.

\item E-Step: Obtain $\{(\widetilde{u}_{1i}, \widetilde{u}_{2i}, \widetilde{u}_{3i}); i = 1, \ldots, n\}$, when $\Theta = \Theta^{(k)}$.

\item M-Step: From ${\cal D}^{(k)} =
\{(\widetilde{u}_{1i}, \widetilde{u}_{2i}, \widetilde{u}_{3i}); i = 1, \ldots, n\}$, using profile likelihood method
obtain $\Theta^{(k+1)}$ which maximizes the complete log-likelihood function based on ${\cal D}^{(k)}$.

\item Check the convergence, if not satisfied then go back to E-Step and continue the process.

\end{enumerate}

\subsection{\sc Testing of Hypotheses}

In this section we discuss different testing of hypotheses problems which have some practical relevance.  We mainly use the
likelihood ratio test (LRT) for the different problems.  In each case under the null hypothesis the MLE of any arbitrary parameter
$\delta$ will be denoted by $\widetilde{\delta}$.

\noindent {\sc Test 1:} We want to test the following hypothesis for unknown $\alpha > 0$
\be
H_0: \alpha_1 = \alpha_2 = \alpha_3 = \alpha \ \ \ \hbox{vs.} \ \ \ H_1: \hbox{At least one is not equal}.
\ee
The problem is of interest as it tests whether all the $U_i$'s have the same distribution or not.  Under $H_0$, the MLEs of
$\alpha$ and $p$ can be obtained using the same EM algorithm as it has been described before.  Here at the M-Step we need
to maximize the function
$$
g(\alpha,p) = g_1(\alpha,p) + g_2(\alpha,p) + g_3(\alpha,p),
$$
with respect  to $\alpha > 0$ and $0 < p < 1$, where $g_1$, $g_2$ and $g_3$ are the same as defined by (\ref{g-1}), (\ref{g-2}) and
(\ref{g-3}), respectively.  Under $H_0$,
\be
2(l(\widehat{\alpha}_1, \widehat{\alpha}_2, \widehat{\alpha}_3, \widehat{p}|{\cal D}) -
l(\widetilde{\alpha}, \widetilde{\alpha}, \widetilde{\alpha}, \widehat{p}|{\cal D})) \longrightarrow \chi^2_2
\ee

\noindent {\sc Test 2:} For $0 < \alpha < 1$, we want to test
\be
H_0: \alpha_1 = \alpha_2 = \alpha, \ \ \alpha_3 = 1 - \alpha  \ \ \ \ \hbox{vs.} \ \ \ \ \ H_1: \alpha_1 \ne \alpha_2.
\ee
This is an important problem as it tests whether the marginals have geometric distribution or not.  We can use the EM algorithm
to compute the MLEs of the unknown parameters under $H_0$.  In this case at the M-Step we need to maximize
$$
g(\alpha,p) = g_1(\alpha,p) + g_2(\alpha,p) + g_3(1-\alpha,p),
$$
with respect to $0 < \alpha < 0$ and $0 < p < 1$, where $g_1$, $g_2$ and $g_3$ are same as defined (\ref{g-1}), (\ref{g-2}) and
(\ref{g-3}), respectively. Under $H_0$
\be
2(l(\widehat{\alpha}_1, \widehat{\alpha}_2, \widehat{\alpha}_3, \widehat{p}|{\cal D}) -
l(\widetilde{\alpha}, \widetilde{\alpha}, 1- \widetilde{\alpha}, \widehat{p}|{\cal D})) \longrightarrow \chi^2_2.
\ee

\noindent {\sc Test 3:} Finally we want to test the following hypothesis
\be
H_0: \alpha_3 = 0 \ \ \ \ \hbox{vs.} \ \ \ \ H_1: \alpha_3 \ne 0.
\ee
This testing problem is of interest as it tests whether the marginals are independent or not.  In this case we do not use the
EM algorithm to compute the MLEs of $\alpha_1$, $\alpha_2$ and $p$ under $H_0$.  However it is an iterative process. We obtain $\widetilde{\alpha}_1$,
$\widetilde{\alpha}_2$ and $\widetilde{p}$ as follows.
\be
\widetilde{p} = \hbox{arg max} \left ( g_1(\widetilde{\alpha}_1(p),p) + g_2(\widetilde{\alpha}_2(p),p) \right ),
\ee
here $g_1(\cdot)$ and $g_2(\cdot)$ are the same defined before, and
$$
\widetilde{\alpha}_1(p) = \hbox{arg max} g_1(\alpha_1,p), \ \ \ \widetilde{\alpha}_2(p) = \hbox{arg max} g_2(\alpha_2,p).
$$
Since $\alpha_3$ is in the boundary, the standard result does not work.  Using Theorem 3 of Self and Liang (1987), it
follows that
\be
2(l(\widehat{\alpha}_1, \widehat{\alpha}_2, \widehat{\alpha}_3, \widehat{p}|{\cal D}) -
l(\widetilde{\alpha}_1, \widetilde{\alpha}_2, 0, \widetilde{p}|{\cal D})) \longrightarrow \frac{1}{2} + \frac{1}{2} \chi^2_1.
\ee

\section{\sc Data Analysis}

In this section we present the analysis of a data set for illustrative purposes.
The data set represents Italian Series A football match score data between `ACF Firontina' ($X_1$) and `Juventus' ($X_2$) during
1996 to 2011.  The data set is presented in Table \ref{football-data}.  It is presented in the contingency table form
in Table \ref{data-cont}.  Before progressing further first we have fitted the DGE to $X_1$, $X_2$ and to max$\{X_1, X_2\}$.  The
MLEs of the unknown parameters and the fitted chi-square values and the associated $p$-values are reported in Table
\ref{one-d-res}.  From Table \ref{one-d-res} it is clear that the DGE fits quite well to $X_1$, $X_2$ and max$\{X_1, X_2\}$.
\begin{table}[h]
\bc
\begin{tabular}{|l|c|c|l||c|c|}  \cline{1-6}

\hline
Obs. & ACF  & Juventus & Obs.   & ACF & Juventus  \\
     & Firontina  &    &    &    Firontina    &           \\
     &   ($X_1$)  & ($X_2$) &  &  ($X_1$)  & ($X_2$)    \\
   &   &   &   &   &   \\   \hline \hline
1 & 1 & 2 & 14 & 1 & 2    \\
2 & 0 & 0 & 15 & 1 & 1    \\
3 & 1 & 1 & 16 & 1 & 3    \\
4 & 2 & 2 & 17 & 3 & 3    \\
5 & 1 & 1 & 18 & 0 & 1    \\
6 & 0 & 1 & 19 & 1 & 1    \\
7 & 1 & 1 & 20 & 1 & 2    \\
8 & 3 & 2 & 21 & 1 & 0    \\
9 & 1 & 1 & 22 & 3 & 0    \\
10 & 2 & 1 & 23 & 1 & 2   \\
11 & 1 & 2 & 24 & 1 & 1   \\
12 & 3 & 3 & 25 & 0 & 1   \\
13 & 0 & 1 & 26 & 0 & 1   \\    \hline
\end{tabular}
\ec
\caption{UEFA Champion's League data \label{football-data}}
\end{table}
\begin{table}[h]
\bc
\begin{tabular}{|c|c|l|c|c|c|}  \cline{1-6}

\hline
$X_1 \downarrow$ $X_2 \rightarrow$ & 0 & 1 & 2 & 3 & Total  \\  \hline
 0 & 1 & 5 & 0 & 0 & 6  \\  \hline
 1 & 1 & 7 & 5 & 1 & 14 \\  \hline
 2 & 0 & 1 & 1 & 0 & 2  \\  \hline
 3 & 1 & 0 & 1 & 2 & 4  \\  \hline
 Total & 3 & 13 & 7 & 3 & 26  \\  \hline
\end{tabular}
\ec
\caption{UEFA Champion's League data \label{data-cont}}
\end{table}
\begin{table}[h]
\bc
\begin{tabular}{|c|c|l|c|c|}  \cline{1-5}

\hline
Data & $\widehat{\alpha}$ & $\widehat{p}$ & $\chi^2$ & $p$-value   \\  \hline
$X_1$ & 4.6681 & 0.2617 & 3.9322 & 0.2689    \\  \hline
$X_2$ & 8.4382 & 0.2311 & 0.0993 & 0.9619   \\  \hline
max$\{X_1, X_2\}$ & 12.2939 & 0.2283 & 1.064 & 0.7857  \\  \hline
\end{tabular}
\ec
\caption{MLEs, chi-square and associated $p$-values for $X_1$, $X_2$ and max$\{X_1, X_2\}$.   \label{one-d-res}}
\end{table}

Now we would like to fit the BDGE model to the above data set.  We use the EM algorithm to compute the MLEs of the unknown
parameters.  We start the EM algorithm with the initial estimates as suggested in the previous section.  If we denote the
initial estimates of $\alpha_1$, $\alpha_2$, $\alpha_3$ and $p$ as $\alpha_1^{(0)}$, $\alpha_2^{(0)}$, $\alpha_3^{(0)}$ and $p^{(0)}$,
respectively, then $\alpha_1^{(0)}$, $\alpha_2^{(0)}$ and $\alpha_3^{(0)}$ satisfy the following equations.
$$
\alpha_1^{(0)} + \alpha_3^{(0)} = 4.6681, \ \ \ \ \alpha_2^{(0)} + \alpha_3^{(0)} = 8.4382 \ \ \ \hbox{and} \ \ \ \
 \alpha_1^{(0)} + \alpha_2^{(0)} + \alpha_3^{(0)} = 12.2939.
$$
Hence,
$$
\alpha_1^{(0)} = 3.8557, \ \ \ \alpha_2^{(0)} = 7.6258, \ \ \ \ \alpha_3^{(0)} = 0.8124.
$$
We have taken $p^{(0)}$ = (0.2617+0.2311+0.2283)/3.0 = 0.2404.  With these initial estimates we have started the EM algorithm and
we stop the EM algorithm when the absolute difference between the two consecutive log-likelihood values is less than $10^{-4}$.
The EM algorithm stops after 24 steps and the final estimates of the unknown parameters and the associated 95\% confidence intervals
are reported within brackets as follows: $\widehat{\alpha}_1$ = 1.2836 (0.5317,2.0355), $\widehat{\alpha}_2$ = 3.7705 (2.0187,5.5223),
$\widehat{\alpha}_3$ = 1.0358 (0.4887,1.5829) and $\widehat{p}$ = 0.3410 (0.2863,0.3957). The associated log-likelihood value becomes
-51.0549.  The expected frequencies for each cell based on the fitted BDGE distribution are provided in Table \ref{exp-data-cont}.
\begin{table}[h]
\bc
\begin{tabular}{|c|c|l|c|c|}  \cline{1-5}

\hline
$X_1 \downarrow$ $X_2 \rightarrow$ & 0 & 1 & 2 & 3   \\  \hline
 0 & 1.28 & 2.60 & 1.43 & 0.57   \\  \hline
 1 & 1.32 & 5.46 & 2.50 & 0.99   \\  \hline
 2 & 0.58 & 1.79 & 2.47 & 0.52   \\  \hline
 3 & 0.21 & 0.66 & 0.43 & 0.78   \\  \hline
 \end{tabular}
\ec
\caption{Expected cell frequencies based on fitted BDGE \label{exp-data-cont}}
\end{table}
The observed chi-square value is 15.8524 with the $p$-value greater than 0.30,
for the $\chi^2$ distribution with 14 degrees of freedom.
Hence, it implies that the BDGE provides a very good fit to the bivariate data set.

It is observed that the final estimates are quite far away from the initial estimates.  So to check, whether they are actually MLEs
or not we have performed a grid search on four dimensions with the range of $\alpha$ values between 0.1 to 10.0 with a grid size
0.0001, and the $p$-values between 0.01 to 0.99 with the same grid size.  We obtain the global optimum values of $\alpha_1$,
$\alpha_2$, $\alpha_3$ and $p$ as 1.2827, 3.7783, 1.0401 and 0.3428, respectively with the log-likelihood value as -51.0538.
Therefore, it is clear that the obtained estimates using EM algorithm are very close to the true MLEs.  It may be mentioned here that
the grid search took more than 6 hours, where as in the same machine the EM algorithm took less than 50 seconds.

Now we would like to test the following hypothesis:
$$
H_0: \alpha_1 = \alpha_2   \ \ \ \ \hbox{vs.} \ \ \ \ \ H_1: \alpha_1 \ne \alpha_2.
$$
It mainly indicates whether the performances of the two teams against each other are the same or not.
Under the null hypothesis we obtain the MLEs as follows: $\widehat{\alpha}_{10} = \widehat{\alpha}_{20}$ = 3.3025,
$\widehat{\alpha}_{30}$ = 1.1423, $\widehat{p}_0$ = 0.3175 and the associated log-likelihood value is -51.9978.  We would like
to use the likelihood ratio test, and the value of the test statistic is 2(-51.0538 + 51.9978) = 1.888.  The associated
$p$-value for $\chi^2_1$ degrees
of freedom is less than 0.17.  Hence, we cannot reject the null hypothesis.
The expected frequencies for each cell based on the fitted BDGE distribution with $\alpha_1 = \alpha_2$, are provided in
Table \ref{exp-data-cont-1}.
\begin{table}[h]
\bc
\begin{tabular}{|c|c|l|c|c|}  \cline{1-5}

\hline
$X_1 \downarrow$ $X_2 \rightarrow$ & 0 & 1 & 2 & 3   \\  \hline
 0 & 1.33 & 1.99 & 0.92 & 0.33   \\  \hline
 1 & 1.99 & 6.06 & 2.23 & 0.79   \\  \hline
 2 & 0.92 & 2.22 & 2.52 & 0.44   \\  \hline
 3 & 0.33 & 0.79 & 0.44 & 0.73   \\  \hline
 \end{tabular}
\ec
\caption{Expected cell frequencies based on fitted BDGE when $\alpha_1 = \alpha_2$.  \label{exp-data-cont-1}}
\end{table}
The observed chi-square value is 18.1072 with the $p$-value 0.20.
Hence, it implies that the BDGE with $\alpha_1 = \alpha_2$ also provides a good fit to the bivariate data set.
Therefore, based on the available data we make the
conclusion that the performances of the two teams against each other are not significantly different.

Now we would like to test whether bivariate geometric distribution can be used or not to analyze this data set, as it has been
considered in Test 2 in the previous section.  We obtain the MLEs of $\alpha$ and $p$ under the null hypothesis
as $\widehat{\alpha}$ = 0.5334 and $\widehat{p}$ = 0.2879.  The associated log-likelihood value becomes -93.3893.
It is observed that the $p$-value of the test statistic, is less than 0.001.  Hence, bivariate geometric distribution cannot
be used in this case.

For comparison purposes we would like to examine whether bivariate Poisson distribution provides a better fit or not to this data set.
We have used the following joint PMF of a bivariate Poisson distribution with parameters $\lambda_1, \lambda_2$ and $\lambda_3$.
\be
P(X_1 = i, X_2 = j) = \sum_{k=0}^{\min\{i,j\}} \frac{e^{-\lambda_1} \lambda_1^{i-k}}{(i-k)!} \times \frac{e^{-\lambda_2} \lambda_2^{j-k}}{(j-k)!}
\times \frac{e^{-\lambda_3} \lambda_3^k}{k!}, \ \ \ \ i,j \in {\cal N}_0.     \label{jpmf-bvp}
\ee
The MLEs of $\lambda_1$, $\lambda_2$ and $\lambda_3$ are as follows: $\widehat{\lambda}_1$ = 0.8089, $\widehat{\lambda}_2$ =
0.9737, $\widehat{\lambda_3}$ = 0.5643, respectively, and the associated log-likelihood value is -53.3251.
The expected frequencies for each cell based on the fitted bivariate Poisson distribution are provided in
Table \ref{exp-data-cont-2}.
\begin{table}[h]
\bc
\begin{tabular}{|c|c|l|c|c|}  \cline{1-5}

\hline
$X_1 \downarrow$ $X_2 \rightarrow$ & 0 & 1 & 2 & 3   \\  \hline
 0 & 2.48 & 2.42 & 1.18 & 0.38   \\  \hline
 1 & 2.01 & 3.36 & 2.32 & 0.98   \\  \hline
 2 & 0.81 & 1.92 & 1.89 & 1.05   \\  \hline
 3 & 0.22 & 0.67 & 0.87 & 0.64   \\  \hline
 \end{tabular}
\ec
\caption{Expected cell frequencies based on fitted bivariate Poisson with joint PMF (\ref{jpmf-bvp}). \label{exp-data-cont-2}}
\end{table}
The observed chi-square value and the associated $p$-value are 21.8381 and 0.08, respectively.
Therefore, it is clear that the bivariate Poisson distribution (\ref{jpmf-bvp}) does not provide a good fit to this data set.

\section{\sc Conclusions}

In this paper we have introduced a bivariate discrete generalized exponential (BDGE) distribution which is a natural discrete version of the continuous
bivariate generalized exponential distribution of Kundu and Gupta \cite{KG:2009}.  The proposed BDGE distribution has the marginals which are discrete
generalized exponential distribution.  The BDGE distribution has four parameters and it is a flexible bivariate model.  We have
derived several properties of the distribution.  It is observed that the MLEs of the BDGE distribution cannot be obtained in explicit forms
and they have to be obtained by solving four non-linear equations simultaneously.  We have proposed a new EM algorithm
which is very specific to this particular problem.  The proposed EM algorithm avoids solving four non-linear equations.  It only needs
to solve four one-dimensional optimization problems.  Hence, it is quite easy to implement in practice.  One data set has been analyzed
and it is observed that the proposed model and the EM algorithm work quite well in practice.

In this paper we have developed mainly the classical inference.  It will be interesting to develop the Bayesian inference also and compare
their performances.  Moreover, here we have considered the bivariate model, it is important to see how it can be generalized to the multivariate case.  More work is needed along these directions.

\section*{\sc Acknowledgements:} The authors would like to thank two unknown referees and the associate editor for their constructive
comments which had helped to improve the manuscript in a significant manner.

\section*{\sc Appendix}

\noindent {PROOF OF RESULT 11:}

First we will show that the function $h(\alpha|p)$ for $0 < p < 1$, and for any $j = 0, 1, \ldots$,
$$
h(\alpha|p) = \ln \left [ (1-p^{j+1})^{\alpha} - (1-p^j)^{\alpha} \right ]
$$
is a log-concave function.  By straight forward calculation it can be seen that
$$
\frac{d^2}{d\alpha^2} h(\alpha|p) = \frac{h_1(\alpha|p)}{h_2(\alpha|p)}<0,
$$
where
\beanno
h_1(\alpha|p) & = & -(1-p^j)^{\alpha}(1-p^{j+1})^{\alpha} \left (\ln(1-p^{j+1})-\ln(1-p^j)\right )^2,   \\
h_2(\alpha|p) & = & \left ((1-p^{j+1})^{\alpha}- (1-p^j)^{\alpha}\right )^2.
\eeanno
\qed

\end{document}